\documentclass{icrc29}
\usepackage{graphicx,amssymb,amsmath,times}
\newcommand{\lepton}{\ifmmode {l} \else $l$\fi}
\newcommand{\elepton}{\ifmmode {l^{*}} \else $l^{*}$\fi}
\newcommand{\wboson}{\ifmmode {{\mathrm W}^{\pm}} \else
${\mathrm W}^{\pm}$\fi}
\newcommand{\wpair}{\ifmmode {{\mathrm W}^{+}{\mathrm W}^{-} } \else
${\mathrm W}^{+}{\mathrm W}^{-}$\fi}
\newcommand{\zboson}{\ifmmode {{\mathrm Z}^{0}} \else
${\mathrm Z}^{0}$\fi}
\begin{document}
\title[Sensitivity of large air shower experiments for new physics searches]{Sensitivity of large air shower experiments for new physics searches}
\author[V.~Cardoso et al.] {V.~Cardoso$^a$,M.C.Esp\'\i rito~Santo$^b$,A.Onofre$^{b,d}$,M.Paulos$^b$,M.Pimenta$^{b,c}$, J.C.Rom\~ao$^c$,B.Tom\'e$^b$ \\
(a) CFC, Universidade de Coimbra, P-3004-516 Coimbra, Portugal \\
(b) LIP, Av. Elias Garcia, 14--1, 1000-149 Lisboa Portugal \\
(c) IST, Av. Rovisco Pais, 1049-001 Lisboa, Portugal \\
(d) Universidade Cat\'olica, Figueira da Foz, Portugal \\
        }
\presenter{Presenter: B.~Tom\'e (bernardo@lip.pt), \  
por-tome-B-abs1-he23-oral}

\maketitle

\begin{abstract}

 Searches for physics beyond the Standard Model of particle physics are 
 performed at accelerators worldwide. Although having poorer detection 
 capabilities and large beam uncertainties, ultra high energy cosmic 
 ray (UHECR) experiments present a unique opportunity to look for new physics 
 far beyond the TeV. Nearly horizontal energetic neutrinos, seeing a large 
 atmospheric target volume and with negligible background from "ordinary" 
 cosmic rays, are ideal to explore rare processes.
 The sensitivity of present and planned experiments to different new 
 physics scenarios is estimated, including mini black-holes, excited 
 leptons and leptoquarks. 
\end{abstract}
\section{Introduction}
Cosmic ray experiments present a unique opportunity to look for new physics at scales 
far beyond the TeV. 
These experiments, covering huge detection areas, are able to explore the 
high energy tail of the cosmic ray spectrum, reaching centre-of-mass 
energies orders of magnitude above those of man made accelerators. 
%
%
Energetic cosmic particles interact with the atmosphere of Earth originating 
Extensive Air Showers (EAS) containing billions of particles. While cosmic particles 
with strong or electromagnetic charges are absorbed in the first layers of the 
atmosphere, neutrinos have a much lower interaction cross-section and can easily 
travel large distances. 
Energetic cosmic neutrinos, although not yet observed and with very large 
uncertainties on the expected fluxes, are predicted on rather solid grounds~\cite{cosmneut}.
Nearly horizontal neutrinos, seeing a large target volume 
and with negligible background from ``ordinary'' cosmic rays, are thus an ideal beam 
to explore possible rare processes~\cite{exoticnu}.
 With large extra dimensions in our universe, black holes (BH) could be produced 
 in UHECR atmospheric interactions. Events with a double bang topology, where 
 the production and decay of a microscopic BH (first bang) is 
 followed, at measurable distance, by the decay of an energetic tau lepton 
 (second bang) could be an almost background free signature. 
 Compositeness is a never discarded hypothesis for explaining the complexity 
 of the fundamental particle picture; leptoquarks arise naturally in models
 unifying the quark and lepton sectors. Excited leptons and leptoquarks could 
 be produced in interactions of quasi-horizontal cosmic neutrinos with the 
 atmosphere, originating detectable air showers.

%

The capabilities of  current (AGASA~\cite{agasa}, 
Fly's Eye~\cite{fly}) and 
future (Auger~\cite{auger}, EUSO~\cite{euso}, OWL~\cite{owl}) very high energy cosmic ray 
experiments to detect these new physics phenomena are discussed. 

\section{Microscopic black hole detection -- the double bang signature}
In the proposed scenario energetic neutrinos ($E_{\nu} \sim 10^6 - 10^{12}$
GeV) interact deeply in the atmosphere (cross-section $\sim 10^3 - 10^7 $ 
pb)
producing microscopic BH with a mass of the order of the neutrino-parton
center-of-mass energy ($\sqrt{s} \sim 1 - 10 $ TeV). The rest lifetime of these BH
is so small ($\tau \sim 10^{-27}$~s) that an instantaneous thermal and democratic
decay can be assumed. The average decay multiplicity ($<N>$) is a function of the parameters 
of the model (Planck mass {$M_D$}, BH mass $M_{BH}$, number of extra dimension $n$ ) 
and typical values of the order of 5-20 are obtained in large regions of the parameter space.
A large fraction of the decay products are hadrons ($\sim 75\%$) but there is a non
negligible number of charged leptons ($\sim 10 \%$)~\cite{halzen,Atlas}. 
The energy spectra of such leptons in the BH centre-of-mass reference 
frame peaks around $M_{BH}/N$. 

Tau leptons provide a ``golden'' signature for microscopic BH detection
in horizontal air shower events~\cite{bh-our}. In fact, in the relevant energy range, the 
tau interaction length in air is much higher than its decay length, which is
given by $L_{decay} = 4.9$ Km ($E_{\tau}/10^8$ GeV)~\cite{halzen}.
A detectable second bang can be produced for tau leptons with a decay
length large enough for the two bangs to be well separated, but small enough 
for a reasonable percentage of decays to occur within the field of view.
Another critical aspect for the detectability of the second bang is 
the visible energy in the tau decay, since a fraction of the energy escapes detection due 
to the presence of neutrinos. In addition, only  decays into hadrons or electrons
originate extensive air showers, leading to observable fluorescence signals.
However, the energy threshold for this second shower
is only determined by the expected number of signal and background photons
in a very restricted region of the field of view, as the second shower
must be aligned with the direction of the first one.

Double bang events in EUSO were generated parameterising the shower development
and the atmosphere response as detailed in~\cite{bh-our}.
The modified frequentist likelihood ratio method~\cite{MFLR},
which takes into account not only the total number of expected signal and background
events but also the shapes of the distributions, was used to compute the statistical 
significance of the second shower. 
The number of background photons has been estimated considering an expected background
rate of 300-500 photons/(m$^2$.ns.sr)~\cite{baby} 
An ideal photon detection efficiency of 1.0 and a more realistic one of 0.1 were considered.

The fraction of the BH events with a first bang within the EUSO
field of view that also have an observable second shower 
is shown in fig.~\ref{fig:results}(a), as a function of $x=M_{BH}/M_D$, for $E_{\nu}=10^{20}$~eV 
and for detector efficiencies of 1.0 and 0.1. These results take into account
the fraction of events with taus in BH decays,
the tau energy spectrum and its decay length, the geometrical
acceptance of EUSO and the visibility of the second shower.


\section{Sensitivity for excited lepton and leptoquark detection}
In models with substructure in the fermionic sector, excited fermion states are 
expected~\cite{hagiboud}.
Excited leptons could be produced in neutrino-parton collisions
via neutral (NC) and charged current (CC) processes, 
$\nu N \rightarrow \nu^* X$ and $\nu N \rightarrow \ell^* X$
($\nu^*$ and  $\ell^*$ representing neutral and charged excited leptons, respectively). 
The hadronic component $X$, and possibly part of the excited lepton decay products, 
would originate an extensive air shower, observable by large cosmic ray experiments.
The strength of the coupling between excited leptons and the SM leptons is parameterised through 
the weight factors $f$ and $f'$  
and the compositeness scale parameter, $\Lambda$. 
The total CC and  NC production cross-sections were computed from the neutrino-parton 
cross-section, as detailed in~\cite{excit-our}. 
For $E_\nu=10^{20}$~eV and $f/\Lambda=15$~TeV$^{-1}$ they range between 50~nb--100~nb 
(between 1~nb--2~nb) for  an excited lepton mass of $m_*=1$~TeV$/c^2$ ($m_*=100$~TeV$/c^2$).
Excited leptons are assumed to decay promptly 
by radiating a $\gamma$, \wboson\   or \zboson\ boson. 
For $\Lambda = 1$~TeV and $E<10^{21}$~eV, their decay length is predicted to be less than 
$10^{-4}$~m and, in all the studied scenarios, they decay essentially at the production point.
In cosmic ray air shower experiments, only the excited lepton decay products 
originating hadronic or electromagnetic showers will contribute to the 
EAS.
%
High energy taus may produce double bang signatures of the type described above.

Leptoquarks are coloured spin 0 or spin $1/2$ particles which
arise naturally in several models attempting the unification of the quark and lepton 
sectors of the Standard Model (SM) of particle physics~\cite{buchmuller:1986zs}. 
Different leptoquark types are expected, according to their quantum numbers, which 
give rise to different coupling strengths and decay modes, and thus to different 
cross-sections and final states.
If the available energies are high enough, the interaction
of cosmic neutrinos with the atmospheric nuclei should create the ideal
conditions for the production of leptoquarks, with dominance of $s$-channel
resonant production.
The produced leptoquarks are expected to decay promptly into
a quark and a charged or neutral lepton. The branching ratio into the 
charged and neutral decay mode depends on the leptoquark type.

The expected number of observed events was obtained from the computed cross-sections,
assuming the Waxman-Bahcall (WB)~\cite{wb} bound with no z evolution for the incident 
neutrino flux, $E_\nu^2 \frac{d\phi}{dE_\nu} = 10^{-8}$ [GeV/cm$^2$ s sr]. 
The procedure outlined in~\cite{excit-our} was followed to obtain estimations of the 
acceptances and observation times of the different experiments. 
The relation between the shower energy and the primary neutrino energy is process dependent.
In the case of excited lepton production, $\nu N \rightarrow \nu^* X$ or 
$\nu N \rightarrow \ell^* X$, 
it depends on the decay mode of the produced neutral or charged excited lepton.
%
%
For each scenario an average acceptance as a function of the incident neutrino energy, 
was computed via Monte Carlo taking into account the $d\sigma_{\nu N}/dy$ distributions 
and the different possible decay modes~\cite{excit-our}.

%

\begin{figure}[hbtp]
\begin{center}
\setlength{\unitlength}{0.0105in}
\includegraphics[width=0.499\linewidth]{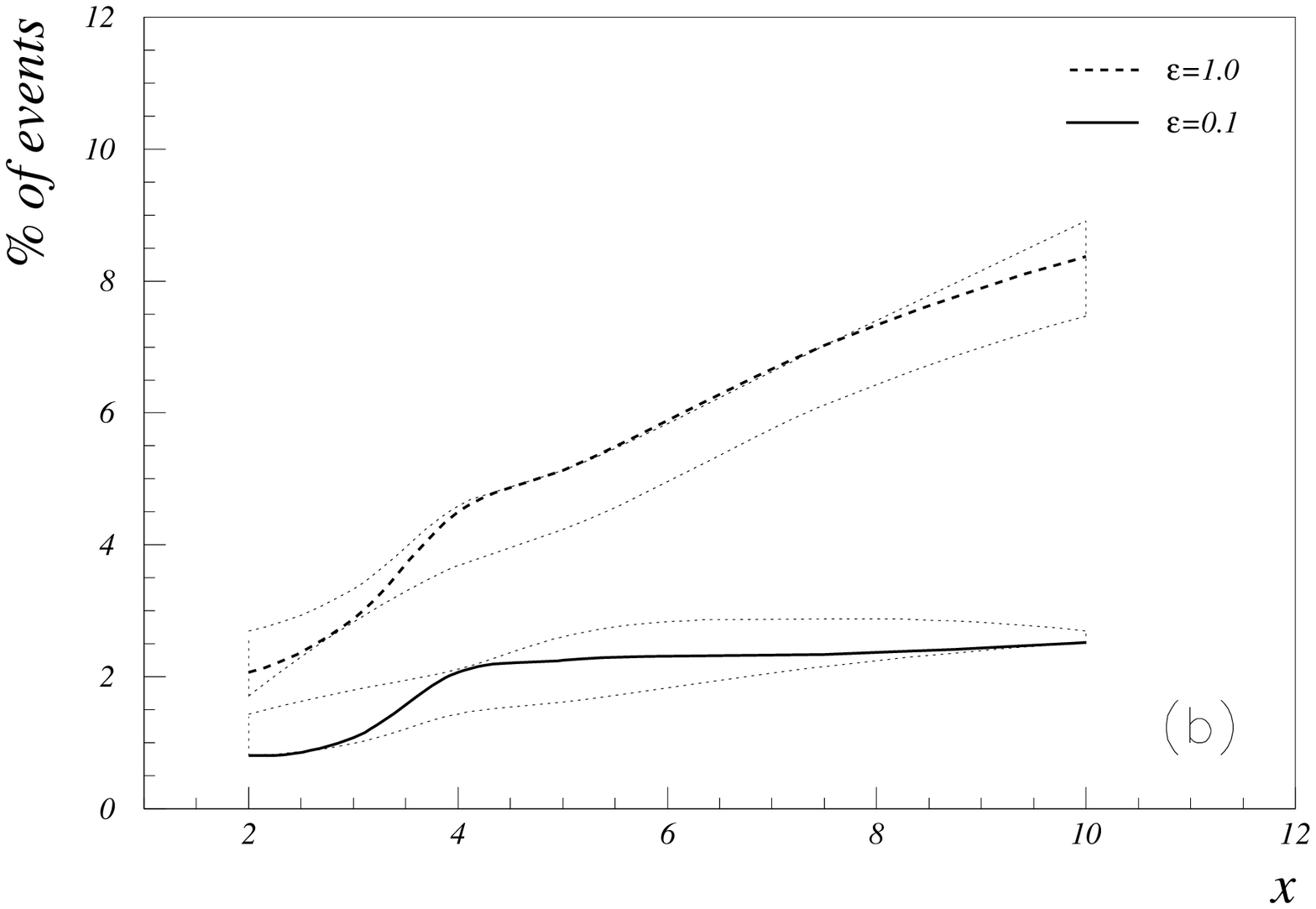}
\includegraphics[width=0.380\linewidth]{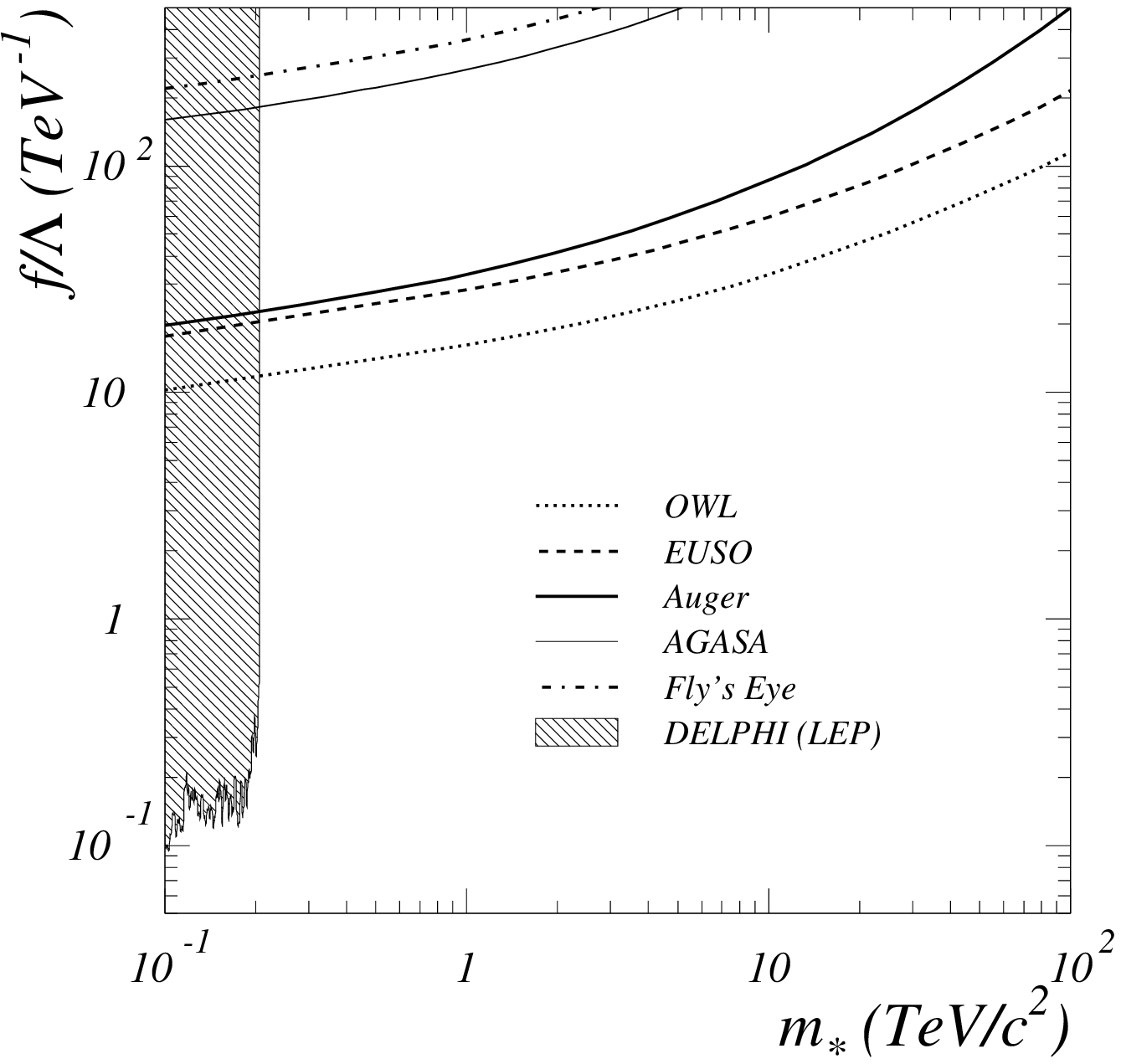}
\end{center}  
\vspace{-0.5cm} 
\caption{a)   
Fraction of the black hole events with a first bang within the EUSO field of view that 
also have a visible second bang,
as a function of $x=M_{BH}/M_D$, for E=$10^{20}$~eV and detection efficiencies $\epsilon=$0.1 
and $\epsilon=$1. The thick lines correspond to $M_D$=1 TeV,$n=3$ and the 
dotted bands give the variation of the results when varying $M_D$ between 1~TeV and 
2~TeV and $n$ between 3 and 6.
b) Estimated sensitivities of the different
experiments as a function of the excited lepton
mass, for excited electrons in the $f=f'$ scenario. 
The regions excluded by the DELPHI experiment at LEP
are also shown (in dashed) for comparison~\cite{lep}.
}
\label{fig:results}
\end{figure}

The sensitivity of the different experiments to excited lepton production, 
as a function of the excited lepton mass, was studied. 
Requiring the observation of one event, the sensitivity on the 
ratio $f/\Lambda$ for excited leptons was derived.
Fig.~\ref{fig:results}(b) shows the obtained 
sensitivities for excited electrons, 
as a function of the excited lepton mass, in the scenario $f=f'$.
The sensitivities for the other excited lepton flavours are comparable but
slightly worse, due to the lower shower energy, for the same energy
of the incident neutrino. 
Sensitivity curves for leptoquarks will soon be available\cite{lq-our}.

\vspace{-0.5cm}

\vspace{-0.5cm}
\section{Conclusions}
Cosmic ray air shower experiments, having access to energy domains far
beyond those of man made accelerators, 
may, in a near future, detect new physics phenomena in interactions of
nearly horizontal energetic neutrinos with the atmosphere.
Events with a double bang topology, an almost background
free signature, have a high discovery potential.
This signature was explored in the framework of the production
of microscopic black holes in the interaction of UHECR in the atmosphere.
The possibility of detecting excited leptons or leptoquarks was also
addressed. 
Excited leptons in a mass range well beyond the TeV scale, could be detected 
if the coupling $f/\Lambda$ is of the order of some tens of TeV$^{-1}$.

\section{Acknowledgements}
V.Cardoso, M.C.Esp\'\i rito~Santo and B.Tom\'e were partially supported through FCT grants 
SFRH/BPD/14483/2003,  SFRH/BPD/5577/2001 and SFRH/BPD/11547/2002.

\end{document}